\documentstyle[osa,manuscript,epsf]{revtex}

\begin{document}

\def\be{\begin{equation}}
\def\ee{\end{equation}}
\def\bea{\begin{eqnarray}}
\def\eea{\end{eqnarray}}
\def\Ui{\:\raisebox{-1.7ex}{$\stackrel{\textstyle U}{i}$}\:}

\begin{flushright}
SU-4240-731\\TIFR/TH/02-06
\end{flushright}
\medskip

\begin{center}
{\Large{\bf Continuous Time-Dependent Measurements: 
Quantum Anti-Zeno Paradox with Applications}} \\[.25in]
{\large A.P. Balachandran$^\dagger$ and S.M. Roy$^{\ddagger}$} \\[1cm]

$^\dagger$ {\it Department of Physics, Syracuse University,}
{\it Syracuse, N.Y. 13244-1130, U.S.A.}\\
E-mail: bal@phy.syr.edu \\

\bigskip

$^{\ddagger}$ {\it Department of Theoretical Physics, Tata
Institute of Fundamental Research}, \\ {\it Homi Bhabha Road, Mumbai 400 005,
India}. \\
E-mail: shasanka@theory.tifr.res.in\\

\vspace{2em}

\end{center}
\bigskip

\begin{abstract} We derive differential equations for the modified
Feynman propagator and for the density operator describing
time-dependent measurements or histories continuous in time.  We
obtain an exact series solution and discuss its applications.  Suppose
the system is initially in a state with density operator $\rho(0)$ and
the projection operator $E(t) = U(t) E U^\dagger(t)$ is measured
continuously from $t = 0$ to $T$, where $E$ is a projector obeying
$E\rho(0) E = \rho(0)$ and $U(t)$ a unitary operator obeying $U(0) =
1$ and some smoothness conditions in $t$.  Then the probability of
always finding $E(t) = 1$ from $t = 0$ to $T$ is unity.  Generically
$E(T) \neq E$ and the watched system is sure to change its state,
which is the anti-Zeno paradox noted by us recently.  Our results
valid for projectors of arbitrary rank generalize those obtained by
Anandan and Aharonov for projectors of unit rank.
\end{abstract}

\vspace{1cm}


\section{INTRODUCTION}

Quantum
physics specifies probabilities of ideal observations at one instant
of time or of a sequence of such observations at different
instants$^1$.  How should one describe the limit of infinitely
frequent measurements or continuous observation?  One of the earliest
approaches to continuous quantum measurements was already suggested by
Feynman$^2$ in his original work on the path integral.  The Feynman
propagator as modified by measurements is to be calculated by
restricting the paths to cross (or not to cross) certain spacetime
regions (where space can mean configuration space or phase space).  An
approximate way of doing this by incorporating Gaussian cut-offs in
the phase space path integral was developed by Mensky$^3$ who also
showed its equivalence to the phenomenological master equation approach
for open quantum systems using models of system-environment coupling
developed by Joos and Zeh and others$^4$.

On the other hand a completely different approach was initiated by
Misra and Sudarshan$^5$ who asked: what is the rigorous quantum
description of ideal continuous measurement of a projector $E$
(time-independent in the Schr\"odinger representation) over a time
interval $[0,T]$?  Their original motivation$^5$: ``there does not seem to be
any principle, internal to quantum theory, that forbids the duration
of a single measurement or the dead time between successive
measurements from being arbitrarily small'', led them to rigorous
confirmation of a seemingly paradoxical conclusion noted earlier$^6$.
The conclusion ``that an unstable particle which is continuously
observed to see whether it decays will never be found to decay'' or
that a ``watched pot never boils''$^7$ was christened ``Zeno's paradox
in quantum theory'' by Misra and Sudarshan$^5$.  The paradox has been
theoretically scrutinized questioning the consistency of infinitely
frequent measurements with time-energy and position-momentum
uncertainty principles$^8$.  Experimental tests$^9$ and their
different interpretations have been rigorously discussed.  

In our recent letter$^{10}$, we showed that in contrast to the
continuous measurement of a time independent projection operator which
prevents the quantum state from changing (the quantum Zeno paradox),
the generic continuous measurement of a time-dependent projection
operator $E_s(t)$ forces the quantum state to change with time (the
quantum anti-Zeno paradox).  We have emphasized that though the two
effects (one inhibiting change of state and the other ensuring change
of state) are physically opposite, they are mutually consistent as
they refer to different experimental arrangements.  We derived the
anti-Zeno paradox in a very broad framework with arbitrary
Hamiltonian, arbitrary density matrix states, and measurement of 
smooth time-dependent projection operators of arbitrary rank.  Our
results are generalisations to projectors of arbitrary rank of the
earlier elegant results for rank one projectors obtained by Anandan
and Aharonov and Facchi et. al.$^{11}$ who considered quantum systems guided
through a closed loop in Hilbert space by measurements represented by
rank one projectors.  They are also generalizations to arbitrary
Hamiltonians of Von Neumann's results on continuous measurements$^1$ in
the case of zero Hamiltonian, and analogous results of Aharonov and
Vardi$^{12}$.  However our results for time-dependent projectors have
a completely different physical origin from those of Kofman and
Kurizki$^{13}$ for time-independent measurements. They showed that
when the frequency of measurements is smaller than a characteristic
difference of eigenfrequencies of the system, an enhancement of decay
can result.  

We ask a question far more general than that of Misra and
Sudarshan: what is the operator (the modified Feynman propagator)
corresponding to an ideal continuous measurement of a projection
operator $E_s(t)$ which has an arbitrary (but smooth) dependence on
time in the Schr\"odinger representation?  We obtain a differential
equation for the operator and a series solution.  We work out several
applications.  One of them leads us to a new watched-kettle paradox
which is apparently quite the opposite of the Zeno paradox, but
mathematically a far reaching generalization of it.  Suppose we
continuously measure from $t = 0$ to $T$ the projector $E_s(t) = U(t)
E U^\dagger(t)$ where $U(t)$ is a unitary operator obeying $U(0) = 1$
and some smoothness conditions, and $E$ a projector obeying $E
\rho(0)E = \rho(0)$, where $\rho(0)$ is the initial density operator.
Then the probability of always finding $E_s(t) = 1$ from $t = 0$ to
$T$ is unity. For the Misra-Sudarshan case, $U(t) = 1$ and we recover the
usual Zeno paradox that the watched kettle does not boil.  Generically
$U(t)$ does not commute with $E$. Hence, for most ways of watching, the 
watched kettle is sure to change its state, an anti-Zeno paradox. If the system
is in an eigenstate of $E$ with eigenvalue unity at $t=0$, it will
change its state with time so as to be in an eigenstate of $E_s(t)$
with eigenvalue unity at all future times.

Our computation of modified Feynman propagators corresponding to
continuous measurements is in the framework of ordinary quantum
mechanics. Exactly the same mathematical expressions for the
propagators would arise in
the `consistent histories' or `sum over histories' quantum mechanics
of closed systems$^{14,15}$, where there is no notion of
measurement. Our computations can therefore be applied also to these
history-extended quantum mechanics provided that probabilities of
measurement outcomes are replaced by weights of histories; the
probability interpretation is restored when the probability sum
rules corresponding to consistency or decoherence
conditions are obeyed.

\section{BASIC PRINCIPLES}

For a
quantum system with a self-adjoint Hamiltonian $H$, an initial 
state vector $|\psi(0)\rangle$ evolves to a state vector $|\psi(t)\rangle$,
$$
|\psi(t)\rangle = \exp(-iHt)|\psi(0)\rangle.
\eqno(2.1)
$$
More generally, an initial state with density operator $\rho(0)$ has
the Schr\"odinger time evolution
$$
\rho(t) = \exp(-iHt) \rho(0) \exp(iHt),
\eqno(2.2)
$$
which preserves the normalization condition ${\rm Tr} ~\rho(t) = 1$.
In an ideal instantaneous measurement of a self-adjoint projection
operator $E$, the probability of finding $E = 1$ is ${\rm Tr} (E\rho E)$
and on finding the value $1$ for $E$, the state collapses according to
$$
\rho \rightarrow \rho' = E\rho E/{\rm Tr} (E\rho E).
\eqno(2.3)
$$
If projectors $E_1,E_2,\cdots,E_n$ are measured at times
$t_1,t_2,\cdots,t_n$ respectively, with Schr\"odinger evolution in
between measurements, the probability $p(h)$ for the sequence of
events $h$, 
$$
h: ~E_1 = 1 ~{\rm at} ~ t = t_1; ~E_2 = 1 ~{\rm at}~ t = t_2; \cdots;
~E_n = 1 ~{\rm at}~ t = t_n
\eqno(2.4)
$$
is$^1$
$$
p(h) = ||\psi_h(t')||^2, ~\psi_h(t') = K_h (t') \psi(0), \ t' > t_n.
\eqno(2.5)
$$
Here $K_h (t')$ is the Feynman propagator modified by the events $h$, 
$$
K_h (t') = \exp(-iHt') A_h(t_n,t_1)
\eqno(2.6)
$$
where,
$$
A_h (t_n,t_1) = E_H(t_n) E_H(t_{n-1}) \cdots E_H(t_1) = T
\prod^n_{i=1} E_H (t_i),
\eqno(2.7)
$$
with $T$ denoting `time-ordering' and the Heisenberg operators
$E_H(t_i)$ are related to the Schr\"odinger operators by the usual
relation 
$$
E_H (t_i) = \exp(iHt_i) E_s(t_i) \exp(-iHt_i), \ E_s(t_i) \equiv E_i.
\eqno(2.8)
$$
The state vector of the system at a time $t'$ after the events $h$ is
$$
\psi_h (t') /||\psi_h(t')||.
\eqno(2.9)
$$
(We shall omit the ket symbol except when confusion can arise
thereby).  Correspondingly, if the initial state is a density operator
$\rho(0)$, the probability $p(h)$ for the events $h$ is given by
$$
p(h) = {\rm Tr} ~K_h(t') \rho(0) K^\dagger_h (t') = {\rm Tr} ~A_h
(t_n, t_1) ~\rho(0) A^\dagger_h (t_n, t_1),
\eqno(2.10)
$$
and the state at $t' > t_n$ is
$$
K_h (t') \rho(0) K^\dagger_h (t')/{\rm Tr} ~(K_h(t') \rho(0)
K^\dagger_h (t')).
\eqno(2.11)
$$
In the history extended quantum mechanics of closed systems$^{14,15}$, 
exactly the same expression (2.10) for $p(h)$ is adopted, with
$h$ denoting the history (2.4) without any mention of
measurements, and $p(h)$ being the weight of the history.  The weight
$p(h)$ is rechristened as probability when certain consistency
conditions are obeyed.

\section{REPEATED MEASUREMENTS WITH ZERO HAMILTONIAN}

We recall first von Neumann's$^1$ fundamental work on the change of
state due to measurements alone, ignoring the Hamiltonian evolution
between measurements.  A state vector $|\phi\rangle$ has the
density operator $\rho_\phi$ obeying
$$
\rho_\phi = |\phi\rangle \langle \phi|, ~~~ \rho^2_\phi = \rho_\phi.
\eqno(3.1)
$$
Given any other pure state $|\psi\rangle$, von Neumann constructed a
beautiful demonstration that repetition of a sufficiently large number
of suitable measurements will change $\rho_\phi$ to an ensemble whose
density operator differs from $\rho_\psi$ by an arbitrarily small
amount.  Now since a good definition of entropy $S(\rho)$ of a state
$\rho$
must (by hypothesis) have the property that measurements only increase
it, we need that 
$S(\rho_\psi) -S(\rho_\phi) \geq 0$.  Interchanging the roles
of $\psi$ and $\phi$, we obtain $S(\rho_\phi) - S(\rho_\psi) \geq 0$.
Therefore, 
$$
S(\rho_\phi) = S(\rho_\psi)
\eqno(3.2)
$$
for any two pure states $\phi,\psi$.  This led von Neumann to define
the entropy corresponding to an arbitrary density operator $\rho$ as 
$$
S = -{\rm Tr}~ \rho~ \ell n ~ \rho,
\eqno(3.3)
$$
a complete set of which is zero for any pure state and positive for
any mixture state.  The von Neumann entropy now plays a fundamental
role in providing a quantitative measure of decoherence, for example
in quantum information processing.

We give von Neumann's demonstration of changing an initial $\rho_\phi$
into $\rho_\psi$ by infinitely repeated measurements in the case of
$\phi$ and $\psi$ being orthogonal states.  (This is enough. If they
are not orthogonal we can find a state $\chi$ orthogonal to both $\phi$ and
$\psi$, change from $\rho_\phi$ to $\rho_\chi$, and then from
$\rho_\chi$ to $\rho_\psi$).

If an observable $R$ with a complete set of nondegenerate orthonormal
eigenvectors
$|\phi_n\rangle$ is 
measured on a state with density operator $\rho$, the states with
density operators $|\phi_n\rangle \langle \phi_n|$ are obtained with
probabilities $\langle \phi_n |\rho| \phi_n\rangle$.  A mixed state
$\rho'$ results.
$$
\rho ~{\buildrel R \over \longrightarrow }~ \rho' = \sum_n E_n \rho E_n, ~E_n
= |\phi_n\rangle \langle \phi_n|.
\eqno(3.4)
$$
This result will be used repeatedly to steer $\rho_\phi$ into
$\rho_\psi$.  Let $k$ be a positive integer and $|\psi^{(\nu)}\rangle$, with
$\nu = 0,1,\cdots,k$, be a set of normalized states $(\parallel
|\psi^{(\nu)}\rangle\parallel = 1)$ which interpolate between $|\phi\rangle =
|\psi^{(0)}\rangle$ and $|\psi\rangle = |\psi^{(k)}\rangle$, e.g.
$$
|\psi^{(\nu)}\rangle = \cos\left({\pi\nu \over 2k}\right) |\phi\rangle +
\sin\left({\pi\nu \over 2k}\right) |\psi\rangle. 
\eqno(3.5)
$$
To $|\psi^{(\nu)}\rangle \equiv |\psi^{(\nu)}_1\rangle$, adjoin a set
of orthonormal 
vectors $|\psi^{(\nu)}_2\rangle$, $|\psi^{(\nu)}_3\rangle$, $\cdots$ to
obtain a
complete orthonormal set of eigenvectors of an observable
$R^{(\nu)}$ with the respective eigenvalues $\lambda_1^{(\nu)},
\lambda_2^{(\nu)}, \cdots$ which are all different.  Starting with the
initial density operator $\rho^{(0)} = |\phi\rangle \langle\phi|$,
successively measure $R^{(1)},R^{(2)},\cdots,R^{(k)}$ to obtain a
final density operator $\rho^{(k)}$:
$$
\rho^0 ~{\buildrel R^{(1)} \over \longrightarrow}~ \rho_1 ~{\buildrel
R^{(2)} \over \longrightarrow}~ \rho_2 \cdots ~{\buildrel R^{(k)}
\over \longrightarrow}~ \rho^{(k)}.
\eqno(3.6)
$$
Here $\rho^{(\nu)}$ is obtained from $\rho^{(\nu-1)}$ after
measurement of $R^{(\nu)}$:
$$
\rho^{(\nu-1)} ~{\buildrel R^{(\nu)} \over \longrightarrow}~
\rho^{(\nu)} = \sum_n E_n^{(\nu)} \rho^{(\nu - 1)} E_n^\nu,
\eqno(3.7)
$$
where
$$
E_n^{(\nu)} = |\psi_n^{(\nu)}\rangle \langle\psi_n^{(\nu)}|.
\eqno(3.8)
$$
The crucial step in proving that $\rho^{(k)} \rightarrow \rho_\psi$
for $k \rightarrow \infty$ will be a lower bound on
$$
\langle \psi|\rho^{(k)}|\psi\rangle = \sum_n \langle \psi^{(k)}|
E_n^{(k)} \rho^{(k-1)} E_n^{(k)} |\psi^{(k)}\rangle = \langle
\psi^{(k)} |\rho^{(k-1)}|\psi^{(k)}\rangle.
\eqno(3.9)
$$
A lower bound can be obtained by repeated application of
\bea \hspace{2.7cm}
\langle \psi^{(\nu+1)} |\rho^{(\nu)}|\psi^{(\nu+1)}\rangle &=& \sum_n
\langle \psi^{(\nu+1)}| E_n^{(\nu)} \rho^{(\nu-1)} E_n^{(\nu)}
|\psi^{(\nu+1)}\rangle \nonumber \\[2mm] &\geq& |\langle \psi^{(\nu+1)}
|\psi^{(\nu)}\rangle|^2 \langle\psi^{(\nu)}
|\rho^{(\nu-1)}|\psi^{(\nu)}\rangle, \nonumber \hspace{3cm} (3.10)
\eea
together with
$$
\langle \psi^{(\nu+1)}|\psi^{(\nu)}\rangle = \cos\left({\pi \over
2k}\right), ~\langle \psi^{(1)} |\rho^{(0)}|\psi^{(1)}\rangle = \cos^2
\left({\pi \over 2k}\right).
\eqno(3.11)
$$
Hence,
$$
\langle \psi|\rho^{(k)}|\psi\rangle \geq \left[\cos\left({\pi \over
2k}\right)\right]^{2k} ~{\buildrel \longrightarrow \over {{}_{k \rightarrow
\infty}}}~ 1.
\eqno(3.12)
$$
Since ${\rm Tr}~\rho^{(k)} = 1$ and $\rho^{(k)}$ is a nonnegative
operator, we have 
\[
\rho^{(k)}_{nn} ~{\buildrel \longrightarrow \over {{}_{k \rightarrow
\infty}}}~ \delta_{n1} ~~~~~~~~~({\rm no~sum~over}~n),
\]
and also, for $m \neq n$,
\[
|\rho^{(k)}_{mn}|^2 \leq (\rho^k)_{mm} (\rho^k)_{nn} ~{\buildrel
 \longrightarrow \over {{}_{k \rightarrow \infty}}}~ 0. 
\]
Hence,
$$
\rho^{(k)} ~{\buildrel  \longrightarrow \over {{}_{k \rightarrow
\infty}}}~ |\psi\rangle \langle \psi|.
\eqno(3.13)
$$
This completes von Neumann's demonstration.

\section{CONTINUOUS MEASUREMENTS WITH ARBITRARY HAMILTONIAN}

The purpose now is to obtain an exact operator expression for the
modified Feynman propagator $K_h (t')$ due to infinitely frequent
measurements in some earlier interval of time allowing for arbitrary
Hamiltonian evolution.  We assume that the projection operators
$E_s(t_i)$ measured at time $t_i$ are values at $t_i$ of a projection
valued function $E_s(t)$. We make also the technical assumption that
the corresponding Heisenberg operator $E_H (t)$ is weakly analytic. We
therefore seek to calculate 
$$
K_h(t') = \exp(-iHt')
A_h(t,t_1),
\eqno(4.1)
$$
where
$$
A_h(t,t_1) = \lim_{n\rightarrow\infty} T \prod^n_{i=1} E_H(t_1 +
(t-t_1) (i-1)/(n-1))
\eqno(4.2)
$$
which is the $n\rightarrow\infty$ limit of Eq. (2.7) with a
specific choice of the $t_i$.  Let us also introduce the projectors
$\bar{E_i}$ which are the orthogonal complements of the projectors
$E_i$,
$$
\bar{E_i} = 1 - E_i
\eqno(4.3)
$$
and a sequence of events $\bar h$ complementary to the sequence $h$,
$$
\bar h : \bar{E_1} = 1 ~{\rm at}~ t = t_1; ~\bar{E_2} = 1 ~{\rm at} ~t
= t_2, \cdots, \bar{E_n} = 1 ~{\rm at} ~ t = t_n.
\eqno(4.4)
$$
Corresponding to Eqs. (2.6), (2.7), (4.1), (4.2), we have equations
with $E \rightarrow \bar E$, $h \rightarrow \bar h$.  Thus,
$$
K_{\bar h} (t') = \exp(-iHt') A_{\bar h} (t,t_1),
\eqno(4.5)
$$
$$
A_{\bar h} (t,t_1) = \lim_{n\rightarrow\infty} T \prod^n_{i=1}
\bar{E_H}(t_1 + (t-t_1) (i-1)/(n-1)).
\eqno(4.6)
$$
The special interest in $K_{\bar h} (t')$ is that it is closely
related to the propagator 
$$
K_{h'} (t') \equiv \exp(-iHt') - K_{\bar h} (t') = \exp(-iHt') [1 -
A_{\bar h} (t,t_1)] , \ h' \equiv \Ui E_i ,
\eqno(4.7)
$$
which represents the modified Feynman propagator corresponding to the
union of the events $E_i$, i.e. to at least
one of the events $E_s (t_i) = 1$ occurring, with $t_i$ lying between
$t_1$ and $t$.  Though the
$E_H(t_i)$ are in general not position projectors, we represent them
in Fig. 1 by space regions and hence we represent $A_h (t,t_1)$ which
is a product of the $E_H (t_i)$ at various $t_i$ by a spacetime
region.  This enables us to visualize the propagator $K_{h'}$ 
as corresponding to Feynman paths which intersect the spacetime
region at least once, the propagator $K_{\bar h}$ as corresponding to
paths which do not intersect the spacetime region at all and the
propagator $K_h$ as corresponding to 
paths which stay inside the region $A_h(t,t_1)$ for all times between
$t_1$ and $t$.  Our object is to obtain exact operator expressions for
the propagators $K_h$, $K_{\bar h}$ which are defined by equations
(4.1), (4.5) with $A_h (t, t_1)$ and $A_{\bar h}
(t, t_1)$ being given by the formal
infinite products in Eqs. (4.2) and (4.6). The
operator results we obtain will also provide evaluations of the path
integral formulae for the propagators in history-extended quantum
mechanics$^{14,15}$. 

\section{DIFFERENTIAL EQUATION AND SERIES
SOLUTION FOR OPERATORS REPRESENTING CONTINUOUS MEASUREMENT}

We see from Eqs. (4.1) and (4.5) that the
modifications of the Feynman propagator due to the sequences of events
$h$ and $\bar h$ consist respectively in multiplication by the operators
$A_h(t,t_1)$ and $A_{\bar h}(t,t_1)$.  Thus $A_h(t,t_1) (A_{\bar
h}(t,t_1))$ represents the continuous measurement corresponding to the
sequence of events $h(\bar h)$.  Consider first the operators
$A_h(t_i,t_1), ~A_{\bar h}(t_i,t_1)$ before taking the $n \rightarrow
\infty$ limit, and note the crucial identities
$$
\bar{E_H} (t_i) A_h (t_i,t_1) = 0, \ E_H (t_i) A_{\bar h} (t_i,t_1) = 0
\eqno(5.1)
$$
which follow from $\bar EE = E\bar E = 0$ for any projection operator
$E$.  Note also that
$$
A_h (t_i,t_1) = E_H (t_i) A_h(t_{i-1},t_1), \ A_{\bar h} (t_i,t_1) =
\bar{E_H} (t_i) A_{\bar h} (t_{i-1},t_1) .
\eqno(5.2)
$$
The relation $\left(\bar E_H (t_{i-1})\right)^2 = \bar E_H (t_{i-1})$
implies $A_{\bar h} (t_{i-1}, t_1) = \bar E_H (t_{i-1}) A_{\bar h}
(t_{i-1}, t_1)$. Hence, 
$$
A_{\bar h} (t_i,t_1) - A_{\bar h} (t_{i-1},t_1) =  
(\bar{E_H} (t_i) - \bar{E_H} (t_{i-1})) A_{\bar h} (t_{i-1},t_1) .
\eqno(5.3)
$$
Dividing by
$t_i - t_{i-1} = \delta t$, taking the limit $n \rightarrow \infty$
(i.e., $\delta t \rightarrow 0$) and assuming that $E_H(t)$ is weakly
analytic at $t=0$ we obtain the differential eqn.
$$
{dA_{\bar h} (t,t_1) \over dt} = {d\bar E_H(t) \over dt} A_{\bar h} (t_-,t_1)
\eqno(5.4)
$$
where the argument $t_-$ on the right-hand side indicates that in case
of any ambiguity in defining the operator product on the right, the
argument of $A_{\bar h}$ has to be taken as $t - \epsilon$ with
$\epsilon \rightarrow 0$ from positive values.  We obtain similarly, 
$$
{d A_h(t,t_1) \over dt} = {dE_H(t) \over dt} A_h(t_-,t_1),
\eqno(5.5)
$$
with
$$
{dE_H(t) \over dt} = i[H,E_H(t)] + \exp(iHt) {dE_s(t) \over dt} \exp(-iHt).
\eqno(5.6)
$$
Further $A_{\bar h} (t,t_1),A_h(t,t_1)$ must obey the initial
conditions
$$
A_{\bar h} (t_1,t_1) = \bar{E_H} (t_1), \ A_h(t_1,t_1) = E_H(t_1).
\eqno(5.7)
$$
The measurement differential equations (5.4) and
(5.5) are reminiscent of Schr\"odinger equation for the
time evolution operator except for the fact that the operators $d\bar
E_H/dt$, $dE_H/dt$ are hermitean whereas in Schr\"odinger theory the
antihermitean operator $H/i$ would occur. Using the
initial conditions (5.7), we obtain the explicit solutions,
$$
A_{\bar h}(t,t_1) = T \exp\left(\int^t_{t_1} dt' {d\bar{E_H} (t')
\over dt'}\right) \bar{E_H} (t_1),
\eqno(5.8)
$$
$$
A_h(t,t_1) = T \exp\left(\int^t_{t_1} dt' {dE_H (t') \over dt'}\right)
E_H (t_1),
\eqno(5.9)
$$
where the time-ordered exponential in (5.8) for example has the
series expansion 
$$
T \exp\left(\int^t_{t_1} dt' {d\bar E_H (t') \over dt'}\right) = 1 +
\sum^\infty_{n=1} \int^t_{t_1} dt'_1 \int^{t'_1}_{t_1} dt'_2 \cdots
\int^{t'_{n-1}}_{t_1} dt'_n T \prod^n_{i=1} {d\bar E_H(t'_i) \over dt'_i}.
\eqno(5.10)
$$
We assume that the time-ordered operator products appearing on the
right-hand side exist at least as distributions.  The distributional 
character occurs naturally for operators with continuous spectrum even
when the $\bar E_H(t)$ (or $E_H (t)$) at different times commute, and
implies that the 
series on the right-hand side must be taken as the definition of the
exponential on the left-hand side; we may not do the integral of
$d\bar E_H(t')/dt'$ on the left-hand side.  (This will be clarified in
examples.) Multiplying the expressions (5.8) and
(5.9) for $A_{\bar 
h}(t, t_1)$ and $A_h (t, t_1)$ on the left by $\exp (-iHt')$ then
completes the evaluation of the modified Feynman propagators $K_{\bar
h} (t')$ and $K_h (t)$.  

\section{EXAMPLES}

\begin{enumerate}
\item[{(i)}] {\it Operator With Continuous Spectrum
Commuting with Hamiltonian}
\end{enumerate} 

For a one-dimensional free particle, $H = p^2/(2m)$, consider measuring 
$$
E_s (t') = \int^{\lambda_R (t')}_{\lambda_L(t')} dp |p> \ <p| 
\eqno(6.1)
$$
continuously for $t' \epsilon [t_1, t]$. Since $E_s(t')$ commutes with
$H$, $E_H (t') = E_s (t')$, and  
$$
\bar E_H (t') = \bar E_L (t') + \bar E_R (t') ,
\eqno(6.2)
$$
where
$$
\bar E_L (t') = \int^{\lambda_L(t')}_{-\infty} dp |p> \ <p| , \ \bar
E_R (t') = \int^\infty_{\lambda_R (t')} dp |p> \ <p| .
\eqno(6.3)
$$
We assume that $\lambda_L (t') < \lambda_R (t'')$ for all $t', t''
\epsilon [t_1, t]$, and $<p|q> = \delta (p-q)$. Hence, $\bar E_L (t')
\bar E_R (t'') = 0$ and Eq. (5.10) yields
$$
A_{\bar h} (t, t_1) = A_L (t, t_1) + A_R (t, t_1) ,
\eqno(6.4)
$$
where
$$
A_L (t,t_1) = \left[ 1 + \sum^\infty_{n=1} \int_{t>t'_n>t'_{n-1} \cdots
t'_1>t_1} dt'_1 \ dt'_2 \cdots dt'_n \ T \prod^n_{i=1} {d\bar E_L
(t'_i) \over dt'_i}\right] \bar E_L (t_1)
\eqno(6.5)
$$
and $A_R$ is given by a similar expression with $L \rightarrow R$. The
orthogonality relations between states $|p>$ imply that the integrand
is a product of $\delta$-functions, 
$$
T \prod^n_{i=1} {d\bar E_L (t'_i) \over dt'_i} \bar E_L (t_1) =
\int^{\lambda_L(t_1)}_{-\infty} dp |p> <p| \prod^n_{i=1} \dot\lambda_L
(t'_i) \delta (\lambda_L (t'_i) - p) . 
\eqno(6.6)
$$
The integrals over $t'_1, \cdots, t'_n$ are now easily done. The
$\delta$-functions vanish for $p < \min \lambda_L$, where $\min
\lambda_L$ denotes the minimum value of $\lambda_L (t')$ for $t'
\epsilon [t_1, t]$. Hence,
$$
A_L (t,t_1) = \bar E_L (t_1) + \int^{\lambda_L (t_1)}_{\min \lambda_L}
dp |p> <p| \sum^{N_p}_{n=1} \sum_{\{t'_1, \cdots t'_n\}} sgn
\left(\prod^n_{i=1} \dot\lambda_L (t'_i)\right) ,
\eqno(6.7)
$$
where $N_p$ is the number of values of $t'$ in the interval $[t_1, t]$
for which $\lambda_L (t') = p$, and for each $n$ we sum over all
$n$-tuples $\{t'_1, \cdots t'_n\} \  {\rm such \ that} \ \lambda_L
(t'_1) = \cdots = 
\lambda_L (t'_n) = p$ with $t > t'_n > t'_{n-1} \cdots t'_1 > t_1$. Hence
$$
A_L (t,t_1) = \int^{\min \lambda_L}_{-\infty} dp |p> <p| +
\int^{\lambda_L(t_1)}_{\min\lambda_L} dp |p> <p| \prod^{N_p}_{i=1}
\left(1+sgn \left(\dot\lambda_L (t'_i)\right)\right) .
\eqno(6.8)
$$
Note that for $N_p = 1$, $\dot\lambda_L (t'_1) < 0$, and that for $N_p
\geq 2$, $\dot\lambda_L (t'_i)$ must have opposite signs for
consecutive integers $i$. Hence, 
$$
\prod^{N_p}_{i=1} \left(1 + sgn (\dot\lambda_L (t'_i))\right) = 0 , \
\ {\rm for} \ N_p \geq 1 .
\eqno(6.9)
$$
An entirely similar evaluation gives $A_R (t, t_1)$. Finally, we get 
$$
A_{\bar h} (t,t_1) = \int^{\min\lambda_L}_{-\infty} dp |p> <p| +
\int^\infty_{\max\lambda_R} dp |p> <p| ,
\eqno(6.10)
$$
where $\max\lambda_R$ is the maximum value of $\lambda_R(t')$ for $t'
\epsilon [t_1, t]$. Of course this answer is correct, and it can
easily be deduced directly from the product of projectors in
Eq. (4.6). But we have obtained here a non-trivial test of the
contribution of terms of arbitrary order in the expansion of the
time-ordered exponential in Eq. (5.8).    

\begin{enumerate}
\item[{(ii)}] {\it Continuous Measurement of Spin
Component along Time-Varying Direction $\vec n (t)$} 
\end{enumerate}

For a spin 1/2 particle with Hamiltonian $H = - (1/2) \sigma_y
\alpha$, let the projector
$$
E_s (t) = {1 + \vec\sigma \cdot \vec n (t) \over 2} ,
\eqno(6.11)
$$
be measured continuously, where
$$
\vec n (t) = (\sin \theta (t) , 0 , \cos\theta (t)) 
\eqno(6.12)
$$
with $\theta (0) = 0$. Defining $\epsilon (t) = \theta (t) + \alpha
t$, we deduce that 
$$
E_H (t) = \exp \left[-{i\over 2} \sigma_y \epsilon (t) \right]
{1+\sigma_z \over 2} \exp \left[{i\over 2} \sigma_y \epsilon (t)
\right] ,
\eqno(6.13)
$$
and that the first five terms in the expansion of the time-ordered
exponential in Eq. (5.8) are (for $t_1 = 0$) given by
\bea
T \exp \left(\int^t_0 dt' \ d \bar E_H (t')/dt'\right) &=& 1 - {1\over
2} \left[\sigma_z (\cos \epsilon - 1) + \sigma_x \sin \epsilon \right]
\nonumber \\ [2mm]
&& + {1\over 4} \left[1-\cos\epsilon - i\sigma_y (\epsilon -
\sin\epsilon)\right] - {1\over 8} \bigg[\sigma_z \{\epsilon
\sin\epsilon + 2\cos\epsilon -2\} \nonumber \\ [2mm]
&& + \sigma_x \{2\sin\epsilon - \epsilon (\cos\epsilon + 1)\}\bigg] -
{1\over 16} \bigg[{1\over 2} \epsilon^2 + \epsilon\sin\epsilon +
3(\cos\epsilon - 1) \nonumber \\ [2mm]
&& + i\sigma_y \{2\epsilon + \epsilon\cos\epsilon - 3
\sin\epsilon\}\bigg] + 0 (\epsilon^5) . \hspace{3.5cm} (6.14) \nonumber
\eea
Note that for $t \rightarrow 0$, $\epsilon (t) \equiv \epsilon$ is of
order $t$ and that the successive square brackets are of orders $\epsilon,
\epsilon^2, \epsilon^3, \epsilon^4$ respectively for $\epsilon
\rightarrow 0$. An analogous result has been obtained by Facchi
et. al.$^{11}$ for a specific time dependence of $\epsilon (t)$.  
Eq. (5.8) then gives $A_{\bar h}$ and
the formula (2.10) the probability $p (\bar h)$ which can be tested
experimentally.  

\section{QUANTUM ANTI-ZENO PARADOX}

We recall
first the usual Zeno paradox. Let the initial state be 
$|\psi_0>$ and let the projection operator $|\psi_0 ><\psi_0|$ be
measured at times $t_1, t_2, \cdots t_n$ with $t_j - t_{j-1} = (t_n -
t_1)/(n-1)$ and $t_n = t$, and let $n \rightarrow \infty$. 
Then, the definition (2.7) yields,
\bea\hspace{.5cm}
A_h (t,t_1) &=& \lim_{n\rightarrow \infty} e^{iHt} |\psi_0> <\psi_0|
\exp (-iH (t-t_1)/(n-1)) |\psi_0>^{n-1} <\psi_0| e^{-iHt_1} \nonumber
\\ [2mm]
&=& \exp (i (H - \bar H) t) |\psi_0> <\psi_0| \exp (-i (H - \bar H)
t_1) , \hspace{4cm} (7.1) \nonumber
\eea
where $\bar H$ denotes $<\psi_0|H|\psi_0>$ and we assume that$^{13}$
$<\psi_0| \exp (-iH\tau)|\psi_0>$ is analytic at $\tau = 0$. Our
differential equation also yields exactly this solution for $A_h (t,
t_1)$. Taking $t_1 = 0$, we deduce that the probability $p(h)$ of
finding the system in the initial state at all times upto $t$ is given
by
$$
p(h) = ||K_h (t) |\psi_0> ||^2 = ||\bar e^{i\bar H t} |\psi_0> ||^2 =
1 ,
\eqno(7.2)
$$
which is the Zeno paradox. (The result can also be generalized to the
case of an initial state described by a density operator, and a 
measured projection operator of arbitrary rank leaving the
initial state unaltered, see below.) 
\bigskip

\begin{center} 
{\bf Anti-Zeno Paradox}
\end{center}

\medskip
The above result may suggest
that continuous observation inhibits change of state. Now we prove a
far more general result which shows that a generic continuous
observation actually ensures change of state. Suppose that the initial
state is described by a density operator $\rho (0)$, and we measure
the projection operator
$$
E_s (t') = U (t') E U^\dagger (t') 
\eqno(7.3)
$$
continuously for $t' \epsilon [0, t]$. Here $E$ is an arbitrary
projection operator (which need not even be of finite rank) which
leaves the initial state unaltered,
$$
E \rho (0) E = \rho (0) ,
\eqno(7.4)
$$
and $U (t')$ is a unitary operator which coincides with the identity
operator at $t' = 0$,
$$
U^\dagger (t') U(t') = U (t') U^\dagger (t') = 1\!\!\!1, U (0) =
1\!\!\!1 .
\eqno(7.5)
$$
The Heisenberg operator $E_H (t')$ is then
$$
E_H(t') = V(t') E V^\dagger (t'), \ \ V(t') = e^{iHt'} U(t') .
\eqno(7.6)
$$
Clearly $V(t')$ is also a unitary operator. The definition (2.7) yields,
for $t_1 \geq 0$, 
$$
A_h (t_n, t_1) = V(t_n) (T \prod^{n-1}_{i=1} X (t_i)) V^\dagger
(t_i), \ n \geq 2 
\eqno(7.7)
$$
where
$$
X(t_i) \equiv E V^\dagger (t_{i+1}) V (t_i) E ,
\eqno(7.8)
$$
and $A_h (t_1, t_1) = V(t_1) E V^\dagger (t_1)$. Denoting
$$
Y (t_j) = T \prod^{j-1}_{i=1} X(t_i) , \ j \geq 2; \ Y(t_1) = E, 
\eqno(7.9)
$$
and noting that $E Y(t_{j-1}) = Y(t_{j-1})$, we have
$$
Y(t_j) - Y(t_{j-1}) = E (V^\dagger (t_j) V(t_{j-1}) - 1) E Y(t_{j-1}).
\eqno(7.10)
$$
Taking $t_{j-1} = t', \ t_j = t' + \delta t, \ n \rightarrow \infty$, we have
$\delta t = 0 (1/n)$, and 
$$
E (V^\dagger (t'+\delta t) V(t') - 1) E = \delta t E {dV^\dagger
(t')\over dt'} V(t') E + 0 (\delta t)^2 .
\eqno(7.11)
$$
To derive that the last term on the right-hand side is $0 (\delta
t)^2$ in the weak sense (i.e., for matrix elements between any two
arbitrary state vectors in the Hilbert space), we make the smoothness
assumption that $E (V^\dagger (t'+\tau) V (t') -1) E$ is analytic in
$\tau$ at $\tau = 0$ in the weak sense. (It may  be seen that this
reduces to analyticity of $<\psi_0| \exp (-iH\tau) |\psi_0>$ in the
usual Zeno case$^{16}$). Hence the $n \rightarrow \infty$ limit yields, 
$$
A_h (t, t_1) = V (t) Y (t) V^\dagger (t_1) ,
\eqno(7.12)
$$
where
$$
{dY(t') \over dt'} = E {dV^\dagger (t') \over dt'} V (t') E Y (t') .
\eqno(7.13)
$$
Solving the differential equation, we obtain, 
$$
A_h (t, t_1) = V(t) T \exp (\int^t_{t_1} dt' E {dV^\dagger (t') \over
dt'} V (t') E) E V^\dagger (t_1) .
\eqno(7.14)
$$
It is satisfying to note that this expression indeed solves our basic
differential equation (5.5) as can be verified very easily
by direct substitution.

The most crucial point for deriving the anti-Zeno paradox is that the
operator 
\[
T \exp (\int^t_{t_1} dt' E {dV^\dagger (t') \over dt'} V (t') E)
\equiv W (t, t_1) 
\]
is unitary, because $(dV^\dagger (t')/dt')V(t')$ is anti-hermitian as 
a simple consequence of the unitarity of $V (t')$.
Taking $t_1 = 0$, Eq. (2.10) gives the probability of finding $E_s (t') =
1$ for all $t'$ from $t' = 0$ to $t$ as 
$$
p(h) = {\rm Tr} \left(V(t) W(t,0) E V^\dagger (0) \rho (0) V(0) E
W^\dagger (t,0) V^\dagger (t)\right) = {\rm Tr} \rho (0) = 1 ,
\eqno(7.15)
$$
where we have used $V(0) = 1$, $E\rho (0) E = \rho (0)$, the unitarity
of $V(t)$ and the unitarity of $W(t,0)$. This completes the
demonstration of the anti-Zeno paradox: continuous observation of $E_s
(t) = U(t) E U^\dagger (t)$ with $U(t) \neq 1\!\!\!1$ ensures that the
initial state must change with time such that the probability of
finding $E_s (t) = 1$ at all times during the duration of the
measurement is unity.  

This remarkable result means that during
continuous observation the quantum state (whether pure or represented
by a density matrix) has an effectively unitary evolution!
Explicitly, for initially pure states 
$$
\psi_h (t) = K_h (t) \psi(0), ||\psi_h (t)|| = 1,
\eqno (7.16)
$$
and for initial density matrix states,
$$
\rho_h(t) = K_h(t) \rho(0) K^\dagger_h (t), \ {\rm Tr} \rho_h (t) = 1.
\eqno (7.17)
$$
Our explicit expressions for $K_h(t)$ yield, 
$$
E_s (t) \psi_h(t) = \psi_h(t)
\eqno (7.18)
$$
and the ``effective'' unitary evolution$^{17}$
$$
i {\partial \psi_h (t) \over \partial t} = \left\{E_s (t) H E_s(t) +
i\left[{dE_s (t) \over dt}, E_s(t)\right]\right\} \psi_h(t),
\eqno (7.19)
$$
the operator in the parenthesis on the right-hand side being
Hermitean.

\section{MEASUREMENTS REPRESENTED BY PROJECTORS OF FINITE RANK AND
COMPARISON WITH PREVIOUS WORK}

In order to compare with previous work$^{11}$ and also to bring out
the simplicity of our explicit formulae consider projectors $E$ (and
therefore $E_s(t)$) of finite rank.

\noindent \underbar{Rank one}.  If $E$ is of rank one, then
$$
E = |\psi(0)\rangle \langle\psi(0)|, \ E_s (t) = |\tilde\psi(t)\rangle
\langle\tilde\psi (t)|,
\eqno(8.1)
$$
where
$$
|\tilde\psi(t)\rangle = U(t) |\psi(0)\rangle.
$$
Our formulae yield, taking $t_1 = 0$,
$$
K_h (t) = U(t) |\psi(0)\rangle \langle\psi(0)| \exp\left(i \int^t_0
dt' \phi(t')\right),
\eqno(8.2)
$$
and
$$
|\psi_h(t)\rangle = |\tilde\psi(t)\rangle \exp\left(i \int^t_0 dt'
 \phi(t')\right),
\eqno (8.3)
$$
where
$$
\phi(t') = \langle\tilde\psi(t')| \left(i {\partial \over \partial t'}
- H\right) |\tilde\psi(t')\rangle,
\eqno (8.4)
$$
which is exactly the result obtained by Anandan and Aharonov$^{11}$.

\noindent \underbar{Rank n}.  If $E$ is of rank $n$, then 
$$
E = \sum^n_{\alpha=1} |\alpha\rangle \langle\alpha|, \ E_s(t) =
\sum^n_{\alpha=1} |\tilde\psi_\alpha(t)\rangle \langle
\tilde\psi_\alpha (t)|,
\eqno (8.5)
$$
where,
$$
|\tilde\psi_\alpha (t)\rangle = U(t)|\alpha\rangle.
\eqno (8.6)
$$
We find that
$$
|\psi_h(t)\rangle = U(t) T \exp\left(\int^t_0 dt'
 \sum_{\alpha,\beta=1}^n |\alpha\rangle f_{\alpha\beta} (t')
 \langle\beta|\right) |\psi(0)\rangle
\eqno (8.7)
$$
where $f_{\alpha\beta} (t')$ is the anti-Hermitian matrix,
$$
f_{\alpha\beta} (t') = i \langle \alpha|U^\dagger(t') \left(i
{\partial \over \partial t'} - H\right) U(t')|\beta\rangle
$$
$$
\hspace*{1.2cm} = i \langle \tilde\psi_\alpha (t')|\left(i {\partial
\over \partial t'} - H\right) |\tilde\psi_\beta (t')\rangle.
\eqno (8.8)
$$
Note that the time-ordering instruction is now essential as the
matrices $f(t')$, and $f(t^{\prime\prime})$ with $t' \neq
t^{\prime\prime}$ do not commute.  Eq. (8.7) is thus a non-trivial
generalisation of the Anandan-Aharonov result (8.3).

\section{MATHEMATICAL REMARKS}

The great
generality of the  
present results with respect to the ordinary Zeno paradox$^5$ derives
from the fact that the unitary operator $V(t)$ need not even obey the
semigroup law$^5$ $V(t) V(s) = V(t + s)$ which played a crucial role
in the Misra-Sudarshan proof.  Further, the 
following remarks about the set of pairs $(E, \rho)$ [with $\rho$
a density operator] fulfilling $E\rho E = \rho$ can be made. The first
is that as $E$ and $\rho$ are self-adjoint, this condition is
equivalent to either of the requirements $E\rho = \rho$, or $\rho E =
\rho$. They mean just that $\rho$ is zero on the range of $(1\!\!\!1 -
E)$.
The properties of the pairs $(E, \rho)$ in a finite-dimensional
quantum theory are simple. In that case, the density operators, being
a convex set, are connected and contractible while the connected
components of projectors $E$ consist of all the projectors of the same
rank. Thus for fixed rank $n$ of projectors, the allowed pairs $(E,
\rho)$ form a connected space with the structure of a fibre bundle,
with projectors forming the base and a fibre being a convex set. This
bundle is trivial, the fibres being contractible. 
If the quantum Hilbert space ${\cal H}_{n+k}$ is of dimension $n+k$,
its unitary group $U (n+k) = \{U\}$ acts on $(E, \rho)$ by
conjugation: $E \rightarrow U E U^{-1} , \ \rho \rightarrow U \rho U^{-1}$.
This action is an automorphism of the bundle. Since any two projectors
of the same rank are unitarily related, it is also transitive on the
base. 
The nature of the base follows from this remark. The stability group
of $E$ is $U(n) \times U(k)$ where $U(n)$ and $U(k)$ act as identities
on the range of $(1\!\!\!1 - E)$ and $E$ respectively. Thus the base,
as is well-known, is the Grassmannian$^{18}$ $G_{n,k} (C) = U
(n+k)/[U(n) \times U(k)]$. 
When we pass to quantum physics in infinite dimensions, the space of
connected projectors are determined by orbits of infinite-dimensional
unitary groups, and, in addition, a projector can itself be of
infinite rank. In this manner, general applications of our results
will involve infinite-dimensional Grassmannians (on which there are
excellent reviews$^{19}$). 

\section{CONCLUSION}

It should be stressed that within
standard quantum mechanics and its 
measurement postulates, both the usual Zeno paradox and the anti-Zeno
paradox derived here are theorems. The two paradoxes appear
`paradoxical' and `mutually contradictory' only when we forget Bohr's
insistence that quantum results depend not only on the quantum state, 
but also on the entire disposition of the experimental
apparatus. Indeed the apparatus to measure $E$ and $U(t) E U^\dagger
(t)$ are different.  It would be interesting to analyse how 
these results appear in a quantum theory of closed systems
(including the apparatus) in which there is no notion of
measurements.  It will also be interesting to devise experimental
tests of the anti-Zeno effect along lines used to test the ordinary
Zeno effect$^9$.  

\bigskip

\begin{center} 
{\large{\bf ACKNOWLEDGEMENTS}}  
\end{center}

\medskip
We would like to thank Guy Auberson, Virendra Singh and Rafael Sorkin
for discussions.We were also greatly helped by Lajos Di\'osi who gave us an 
alternative proof of (7.19) and by Jeeva Anandan who drew our attention to 
the first paper of ref 11. Part of this work was supported by the U.S. DOE 
under contract no. DE-FG02-85ER40231, and by the Indo-French Centre For 
Promotion of Advanced Research (IFCPAR/CEFIPRA), project 1501-2.

\newpage

\begin{figure}[htb]
\begin{center}
\leavevmode
\hbox{%
\epsfxsize=6.5in
\epsffile{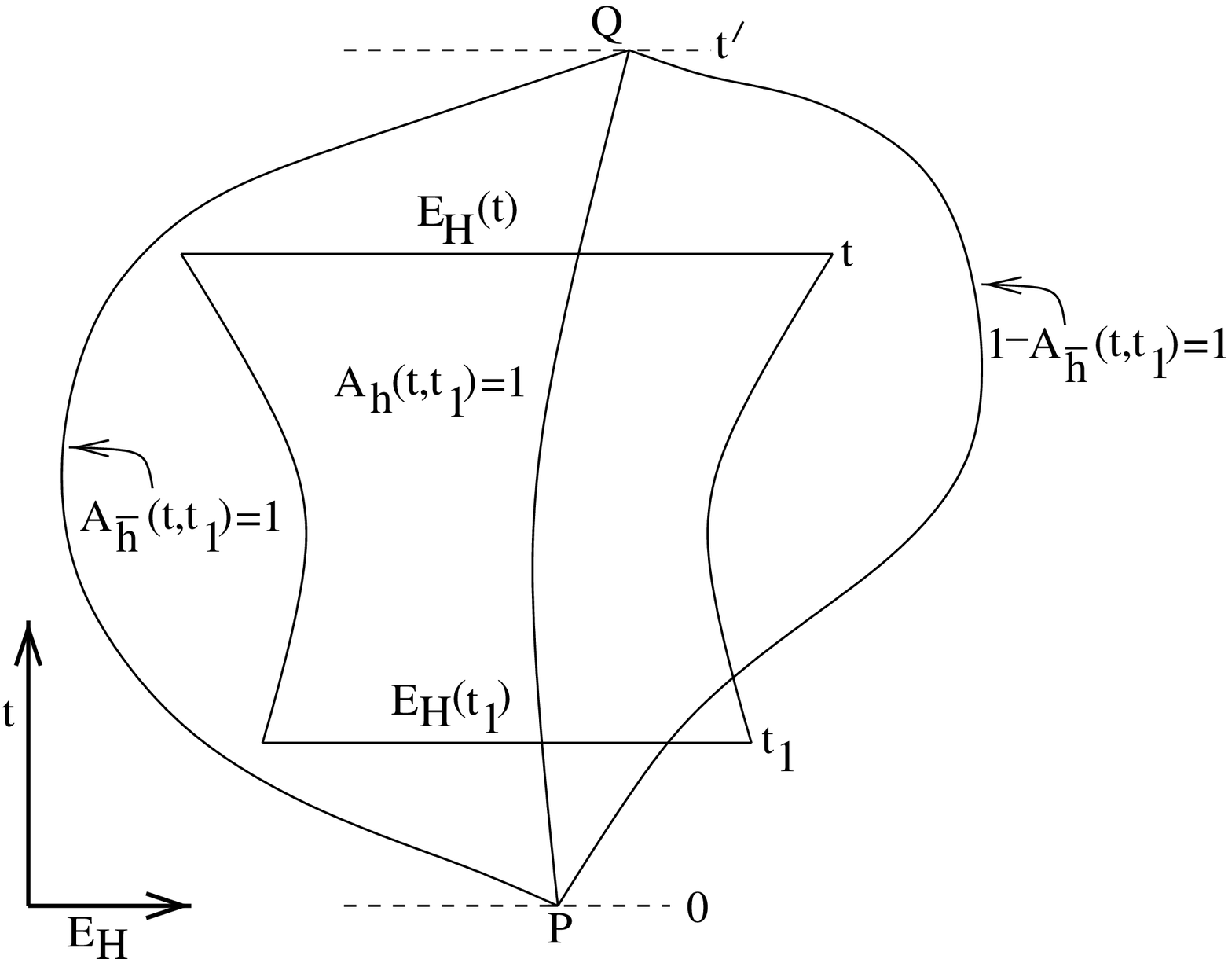}}
\label{fig:9924fig1}
\caption{We may visualize the product of projectors $A_h(t,t_1)$ by a
spacetime region if we represent $E_H(t)$ by a space region (though
$E_H(t)$ need not be a position projector).  In the path integral
approach sum over paths which intersect the spacetime region at least
once $(1 - A_{\bar h} (t,t_1) = 1)$ yield the propagator
$K_{h'} (t')$, paths which stay inside the region for times between
$t_1$ and $t$, $(A_h(t,t_1)=1)$ yield the propagator $K_h(t')$, and paths
which do not intersect the spacetime region at all $(A_{\bar h}
(t,t_1) = 1)$ yield $K_{\bar h} (t')$.}
\end{center}
\end{figure}

\end{document}